\documentclass[12pt]{article}
\usepackage{graphicx}
\newcommand {\bi} {\bibitem}
\newcommand {\be} {\begin{equation}}
\newcommand {\ee} {\end{equation}}
\newcommand {\bea} {\begin{eqnarray} }
\begin{document}

\title{Cooperative behavior in a spatial model of ``commons''}
  \author{Agha 
  Afsar Ali \\ Harish Chandra Research Institute \\ Chhatnag Road,
  Jhunsi, Allahabad-211 019, India\\
  agha@mri.ernet.in} 
\maketitle

\begin{abstract}
  We study a lattice model of ``commons'', where a resource is shared locally
  among the agents of various cooperative tendency. The payoff function of an
  agent is proportional to the fraction of his operation rate and the net
  output of the resource. After each time step a site is occupied by the
  neighbor of maximum profit or by its owner himself. In steady state the
  model is dominated by ``altruist'' agents with a small minority of selfish
  agents, forming a complex pattern. The dynamics selects cooperative levels
  in a way that the model becomes critical. We study the critical behavior of
  the model in case of moderate mutation rate and find the power spectrum of
  fluctuation of activity shows a $1/f^\alpha$ behavior with $\alpha \sim
  1.30$. In case of very slow mutation rate the steady state has slow
  fluctuations which helps the evolution of higher cooperative tendency.
\end{abstract}
PACS numbers:02.50.Le,05.40.-a,05.65.+b

\section{Introduction}

Selfishness is regarded as the rational behavior of agents in various social
and economic studies. However, there are cases where altruistic or cooperative
tendency can survive. The game of ``prisoners dilemma'' is one such case where
cooperative behavior is necessary to maximize ones payoff \cite{Ax}.  Axelrod
and Hamilton have shown that selfish agents with finite memory will eventually
learn to cooperate if the game is played many times \cite{Ax}. In a later
work, May and Nowak showed that in a spatially extended population, altruistic
or cooperative tendency can survive by forming spatial territories\cite{NM}.
A single altruist is not strong but it becomes strong in a group because of
its cooperative tendency.  Thus once a domain of altruists is formed in space
it can survive against the selfish members of the species.  In some biological
systems altruistic behavior is observed instead of purely selfish behavior
\cite{SZQ}.

The problem of ``tragedy of commons'' is the other such case where the selfish
tendency to maximize one's gain leads to a degeneration of the common
resource, and thus reduces everyones gain \cite{H}. We study a spatial version
of this problem in which we assume a resource spread over all the sites of the
lattice.  If an uncontrolled exploitation by a few agents leads to a local
degeneration of the resource; its negative effect is felt only in an immediate
neighborhood.  Similarly, if the resource is maintained in a good condition,
its positive effect is also felt locally. Thus, the resource is shared only
locally. This is the situation observed in case of most natural resources.

\begin{figure}
  \centering
  \includegraphics[width=0.5\textwidth]{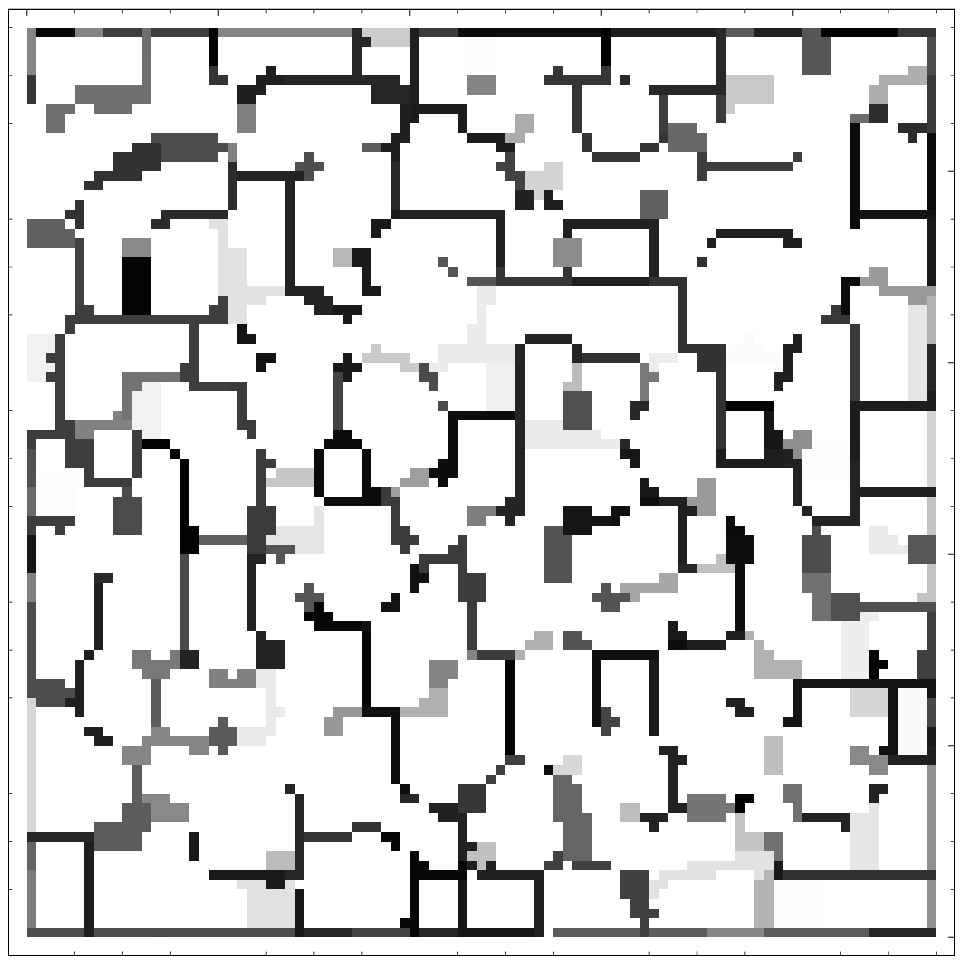}
  \includegraphics[width=0.5\textwidth]{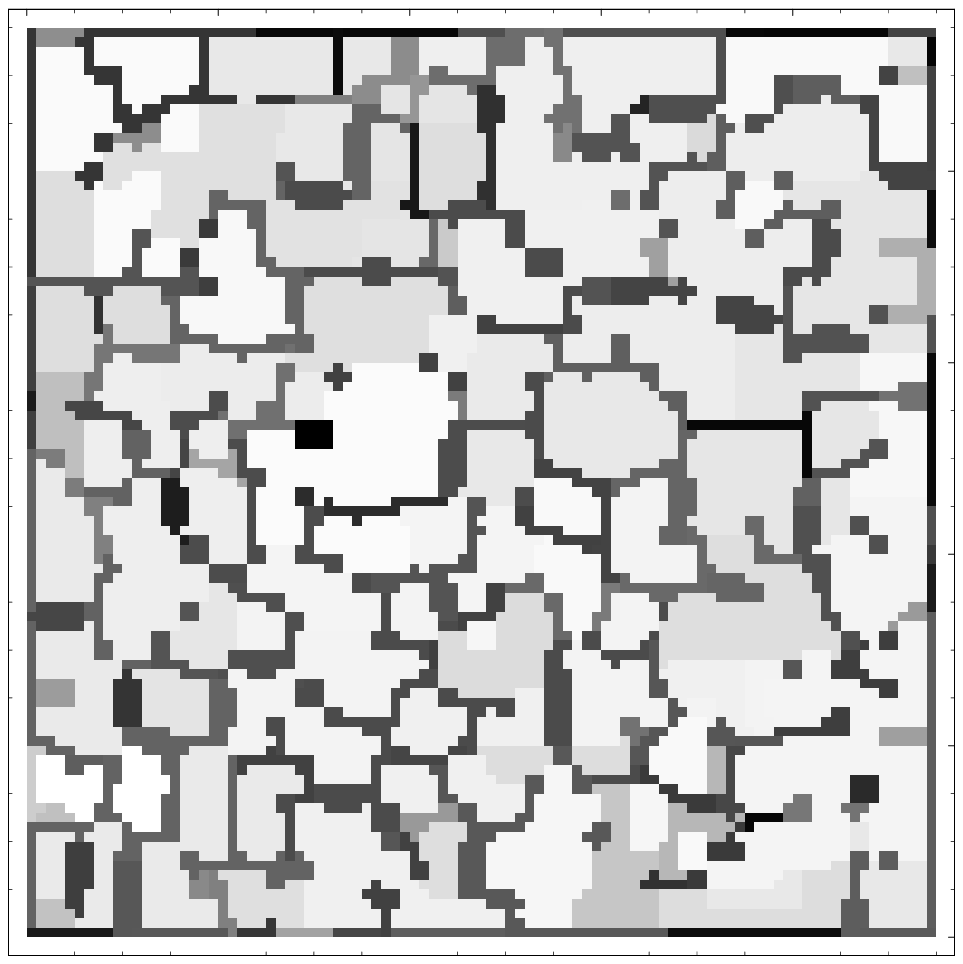}
\caption{Typical spatial pattern for lattice size $L=95$, and mutation rate $p
  =0.001$ (above). The lighter gray shades represent the higher cooperative
  levels and the darker gray shades represent the lower cooperative levels. In
  case of the slow mutation rate, i.e., $p=0.000001$ (below) the cooperative
  band is broad and contains many levels shown by the different gray shades.}
\label{cfs}
\end{figure}

We model this situation with evolution rules similar to those of May and
Nowak.  In our model, instead of only two types of agents, there are many
types of agents whose cooperative tendencies are labeled by a real variable
between $0$ and $1$.  The agents of type $1$ are totally cooperative and the
agents of type $0$ are totally selfish, and the agents of type between $0$ and
$1$, represent a mixed strategy. In the steady state, the density profile of
agents develops two peaks, one near $0$ representing the defector band and the
other near $1$, representing the cooperator band.  Agents from these two bands
form a complex spatial pattern (see Fig.  \ref{cfs}).  Members of the defector
band form thin lines whereas members of the cooperator band form compact
domains, which we will also call colonies. It should be emphasized that the
colonies are formed only by cooperators, demonstrating the advantage of
cooperative over selfish behavior.  This pattern is similar to the one found
by May and Nowak in certain range of the parameter used in their model.

The complex pattern formed in this model has some interesting statistical
properties.  If we allow {\em mutation}, i.e., a change of strategy, of agents
at a moderate rate, then we find that the model self-organizes to a critical
steady state.  The power spectrum of fluctuation of the activity has a
$1/f^\alpha$ form with $\alpha = 1.30$. The critical behavior is lost for very
slow mutation rate.  Thus this case of self-organized criticality is different
from that of the sandpile model which is critical at very slow drive rate
\cite{SOC}.

The moderate mutation rate helps evolution of cooperation in the initial
stage.  But, the large critical fluctuations restrict the cooperative tendency
to a maximum of $8/13$.  At very slow mutation rate an interesting process of
evolution starts. A more stable steady state helps the growth of cooperative
tendency higher than $8/13$.

\begin{figure}
  \centering
  \includegraphics[width=0.5\textwidth]{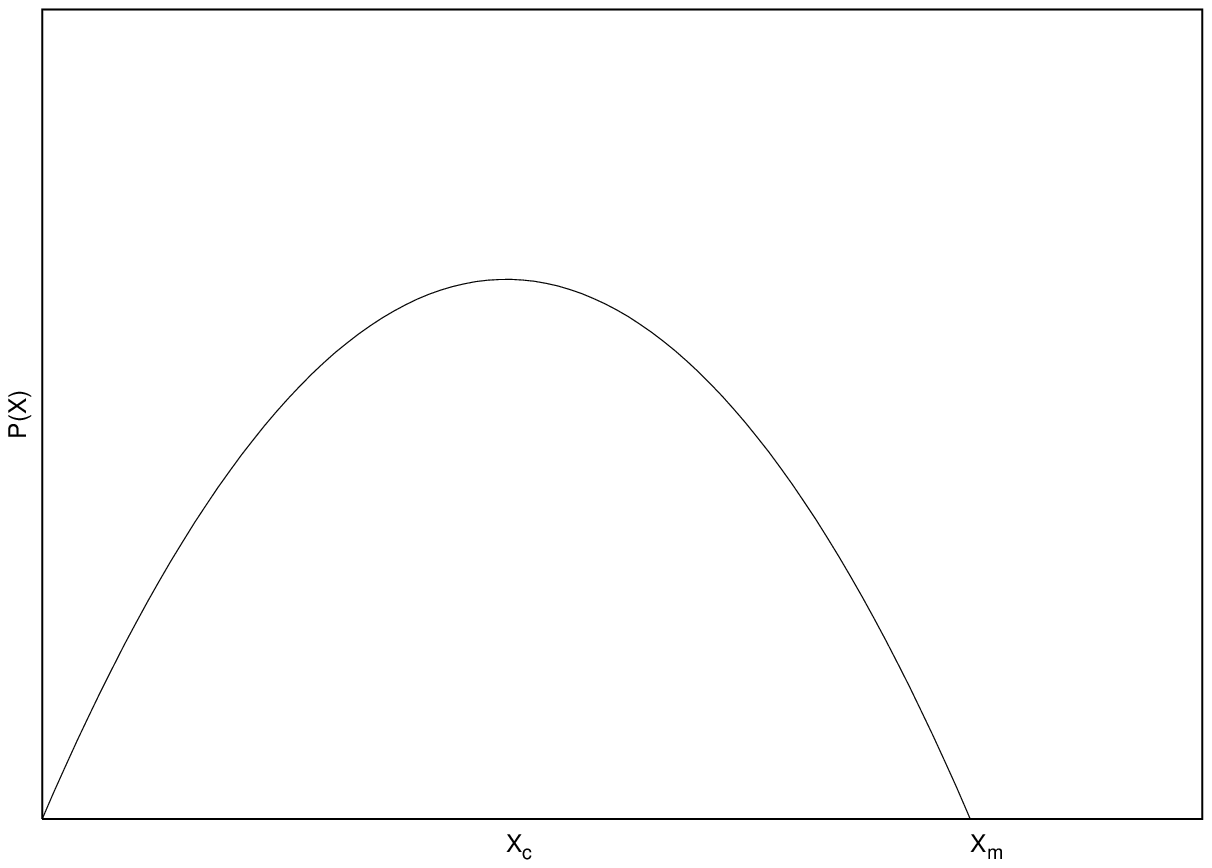}
\caption{The yield function of the resource.}
\label{form}
\end{figure}

\section{``Tragedy of commons''}

Consider a {\em resource} shared by a group of $N$ agents.  When the {\em
  exploitation rate} $X$ is very slow compared to the {\em regeneration rate}
of the resource, the {\em yield} $P(X)$ is proportional to $X$. However, when
$X$ becomes approximately equal to the regeneration rate the yield $P(X)$
reaches a saturation point. If the resource is exploited at an even faster
rate, it degenerates, and $P(X)$ decreases.  Thus the yield $P(X)$, at the
exploitation rate $X$ has the form shown in Fig. \ref{form}.  $P(X)$ is
maximum at $X_c$, and goes to zero at $X_m$ (the resource is fully destroyed
at $X_m$).  For definiteness, we assume the form $P(X)=X(2 -{X\over N})$. So
$X_c=N$, and $X_m=2N$. The income of the $i$th agent $P_i(x_i)$ is
proportional to his exploitation rate $x_i$.

\be 
P_i(x_i)={x_i\over X} P(X)=x_i(2 -{X\over N})\ ,
\label{eq1}
\ee 
where $\sum x_i=X$.

Consider the case when all the agents operate at a rate $x = 1$, the income of
each of them is $1$, and net income of the group is maximum.  Now, let the $j
th$ agent increases his exploitation rate from $1$ to $2$.  Then $X=N+1$,
$P_i=1-1/N$, for $i\ne j$, and $P_j=2(1-1/N)$.  Thus the $j$th agent almost
doubles his income, while the other members incur a small loss of order $1/N$.
The net income of the group also decreases by a fraction of order $1/N^2$.
Therefore, if the agents try to maximize their personal income, the system is
pushed beyond the optimum exploitation rate $X_c$, towards $X_m$, where the
income of all agents goes to zero, and the ``tragedy of commons'' occurs
\cite{H}.

One observes that in this case, there is a conflict between individual
interest and global interest ( and eventually the individual interest has
conflict with itself).  Thus if we try to maximize the income of each agent by
the set of equations of form $\frac {dx_i}{dt}=\Gamma\frac {dP_i(x_i)}{dx_i}
+\eta $, we find that this set of equations, instead of maximizing the
$P_i$'s, lead to a state of lower $P_i$'s.  Hence, the separate optimization
of the individual incomes, does not work.

In the physical systems this kind of a situation can be seen in case of
granular media. If sand is allowed to flow in presence of the gravitational
field, the individual tendency of grains to minimize their potential energy
may lead to a jammed configuration. Thus individual tendency has conflict with
itself.

Here, we try to understand that how cooperation can evolve so that
the ``tragedy of commons'' can be averted.

\section{Model of locally shared resource (LSR)}

We study a model of locally shared resource (LSR). This is defined on a square
lattice, with a real variable $s_i\in[0,1]$, called the cooperative level of
the occupant of the $i$th site. This represents the cooperative tendency of an
agent.  The exploitation rate depends on level $s_i$ through, $x_i=2-s_i$ (
clearly $x_i \in [1,2]$).  The higher $s_i$ implies higher cooperative
tendency and lower $x_i$. The agents of $s<{1\over 2}$, will be called
``defectors'' or ``selfish'', and those of $s \ge {1\over 2}$ will be called
``cooperators'' or ``altruists''.  The resource near the site $i$ is shared in
a neighborhood $D(i)$, which consists of the occupants of the $i$th site and
its nearest and next nearest neighbors. Thus $N=9$ in two dimensions.  The
payoff of the occupant of the $i$th site can be expressed in terms of $s_i$
using (\ref{eq1}) and the relation $x_i=2-s_i$.

\be 
P_i={1\over N}(2-s_i)\sum_{j\in  D(i)} s_j\ . 
\label{eq2}
\ee 

The evolution rule consists of two parts: 

\begin{enumerate}
\item Mutual struggle: After one time step, the site $i$ will be occupied by
  the neighbor of highest payoff, {\it i.e.}, $s_i \rightarrow s_h$, where
  $h\in D(i)$ has highest income $P_h$.
  
\item Mutation: The levels of a small fraction $p$ of agents are changed to
  randomly chosen levels.
\end{enumerate}

We study this model with both periodic and fixed boundary conditions.  In case
of fixed boundary condition, the set $D(i)$ has only four elements for the
sites on the corners and it has six elements for the sites on the edges. In
case of periodic boundary condition, $D(i)$ contains nine elements for all
$i$.

The first rule of evolution in our model is same as that in the model of May
and Nowak. However in their model, there are only two types of agents: One who
always cooperates ($s=1$) and the other who always defects ($s=0$). The game
of prisoners' dilemma is played between two adjacent neighbors. If both
cooperate then the payoff of each of them is 1. If one defects while the other
cooperates then the payoff of the defector is $b > 1$, while the payoff of the
cooperator is 0.  If both defect then the payoff of each of them is 0. If
agent of the $i$th site plays this game with all his eight neighbors, then his
payoff is

\be 
P_i= [b-(b-1)s_i]\sum_{j\in D(i)-i}s_j\ .
\label{eq3} 
\ee 

In the LSR model, there are infinitely many types of agents labeled by a
continuous variable $s$.  If we consider a population of only two levels
$s^1$, and $s^2$, then this model is similar to the model of May and Nowak.
Note that the two models are not identical because the summation in
(\ref{eq2}) for the LSR model includes $i$. The effective $b$ can be defined
as the ratio of payoffs of $s^1$ and $s^2$ in identical neighboring
configurations.  Ignoring $O(1/N)$ correction, we can write
$b(s^1,s^2)\sim(2-s^1)/(2- s^2)$.  However, it is important to point out that
$b$ alone does not determine the behavior of the model.

\begin{figure}
  \centering
  \includegraphics[width=0.7\textwidth]{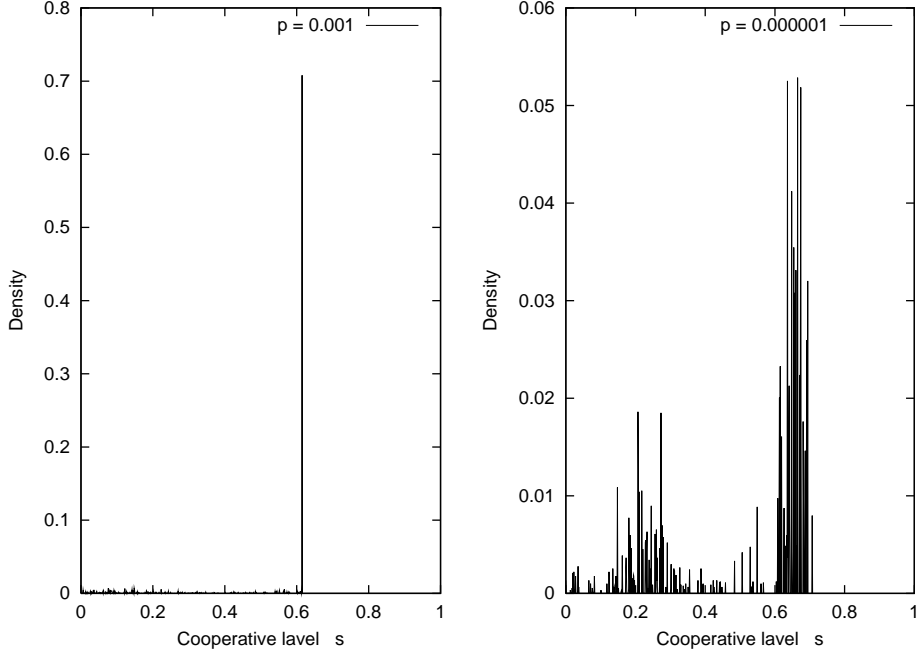}
\caption{The density profile for  the moderate mutation rate (left). There is a
  huge peak at level $s=8/13$. In case of the slow mutation rate (right) the
  levels $s> 8/13$ have also got populated.}
\label{dprofile}
\end{figure}

\begin{figure}
  \centering
  \includegraphics[width=0.7\textwidth]{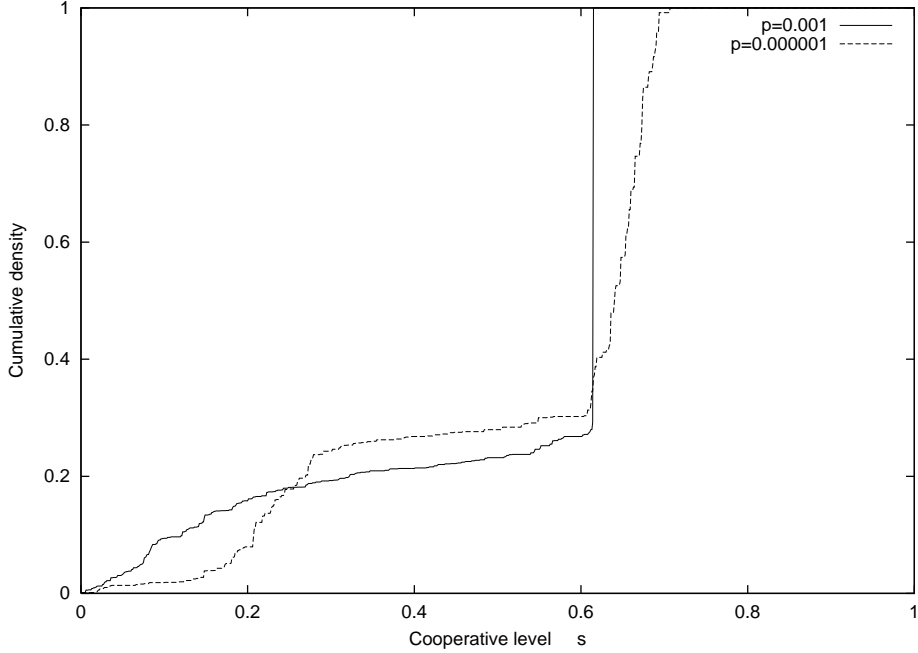}
\caption{The cumulative density showing the fraction of population in
  cooperative levels below level $s$. In case of a moderate mutation rate ($p
  = 0.001$), there is a huge jump at $s=8/13$. Approximately $71\%$ of
  population lies in this level. In case of a very slow mutation rate levels
  $s>8/13$ also get populated.}
\label{cprofile}
\end{figure}

\section{Density profile in LSR model}

We simulate this model for different mutation rates on lattices of various
sizes.  The statistical behavior of the model does not depend on the boundary
condition.  However, it depends on the mutation rate. For moderate mutation,
i.e. $p > 0.0005$ approximately, the model shows critical behavior. In this
case, the average cooperative level $\langle s\rangle \simeq 0.50$. In case of
very slow mutation rate, i.e. $p < 0.0005$ approximately, the model slowly
develops higher cooperative regions, and the average cooperative level rises
to $\langle s\rangle \simeq 0.53$.

In both cases the highest cooperative members (i.e.  $s=1$) do not survive,
but a little practical agents from a band of lower cooperative levels, i.e.
${8\over 13}<s <{5\over 7}$, dominate in the steady state (see the second peak
in the density plot, Fig. \ref{dprofile}, and second step in the cumulative
plot, Fig. \ref{cprofile}.). We call it the {\em cooperator band} (C-band).
This band constitutes around $71\%$ of the population. In case of a moderate
mutation rate this band shrinks to just one level $s={8\over 13}$.  In case of
a slow mutation rate the higher levels also form colonies. As a result the
population of cooperators is evenly distributed in range ${8\over 13} \le s
\le {5\over 7}$.

Around $25\%$ of the population come from the {\em defector band} (D-band),
corresponding to levels, $0.1<s<0.3$ approximately (see the first peak in
density plot, Fig. \ref{dprofile}, and the first step in the cumulative plot,
Fig. \ref{cprofile}.). The level lying between these two bands, i.e.  $0.3 < s
< 0.6 $ approximately, are almost absent from the population.  Thus we find
that in steady state the density profile of cooperative levels develops two
peaks.

The C-band is very sharp, while the D-band is wide and contains more levels.
This is because the spatially close levels have very high mutual competition.
As a result the levels of C-band eliminate each other and in the end only a
few levels survive. The levels of D-band have low density. They exist as
scattered separated lines. A defector of one level hardly encounters a
defector of another level. Therefore, they are not able to eliminate each
other.

\begin{figure}
  \centering
  \includegraphics[width=0.7\textwidth]{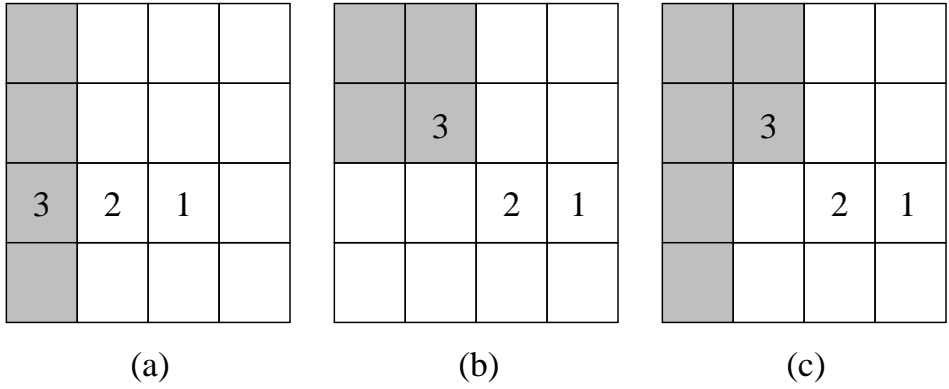}

\caption{The subconfigurations showing (a)edge, (b)corner and (c) step. The
  gray site denoted $s^\prime$, and white site denotes $s$. The full
  configuration can be obtained by extending this pattern.}

\label{ECS}
\end{figure}

A good deal of understanding about the density profile of agents of various
levels can be obtained by considering the stability of some of the special
cases. 

{\bf Edge stability}: Consider the domains of two types of agents $s$ and
$s^\prime$, with $s >s^\prime$, separated along a straight line (see Fig.
\ref{ECS}). The agents of level $s$ is represented by white color, and the
agents of level $s^\prime$ is represented by gray color. The agent $1$ is
among the strongest white agent which can have influence on the domain
boundary. The agent $2$ is the strongest white agent near the domain boundary.
And, the agent $3$ is the strongest gray agent near the domain boundary.  The
payoffs of these three agents are, $P_1=(2-s)s$, $P_2={1\over
  9}(2-s)(6s+3s^\prime)$, and $P_3={1\over 9}(2-s^\prime)(3s+6s^\prime)$.  Now
there are three possibilities:

\begin{itemize}
\item $P_2>P_3$, {\it i.e.} $s^\prime < 1-s$. In this case the white agent $2$
  can take over the gray agent $3$, and the edge will move towards the domain
  of the level $s^\prime$.  Hence, the level $s$ will be called {\em
    edge-strong} against the level $s^\prime$.  The levels $s<{1\over 2}$ are
  edge-strong against all the lower levels.
  
\item $P_1<P_3$, {\it i.e.}  $2-{3\over 2}s<s^\prime<s$. In this case the gray
  agent $3$ can take over his neighboring white agents including agent $2$, and
  the edge will move towards the domain of the level $s$, so the level $s$ is
  {\em edge-weak} against the level $s^\prime$. The levels $s <{4 \over 5}$
  are never edge-weak against any of the lower levels.
  
\item $P_2<P_3$, but $P_1>P_3$, {\it i.e.}  $1-s<s^\prime<2-{3\over 2} s$. In
  this case the edge will be stable, so the level $s$ will be called {\em
    edge-stable} against the level $s^\prime$.
\end{itemize}

{\bf Corner stability}: Consider now the case in which a domain of level
$s^\prime$ forms a corner in a domain of level $s$, and $s>s^\prime$ (see Fig.
\ref{ECS}).  Here again we select three agents using the same criterion as in
the previous case.  The payoffs of these three agents are, $P_1=(2-s)s$,
$P_2={1\over 9}(2-s)(8s+s^\prime)$, and $P_3={1\over
  9}(2-s^\prime)(5s+4s^\prime)$. Again there are three
possibilities:

\begin{itemize}
\item $P_2>P_3$, {\it i.e.} $s^\prime < {1\over 2}(3-4s)$. In this case the
  corner of $s^\prime$ will be decimated. Hence, the level $s$ will be called
  {\em corner-strong} against the level $s^\prime$. The levels $s < {1\over
    2}$ are corner-strong against all the lower levels, and the levels $s> {3
    \over 4}$ are never corner-strong against any of the lower levels .
  
\item $P_1<P_3$, {\it i.e.}  $2-{9\over 4} s<s^\prime<s$. The domain of
  $s^\prime$ will grow near the corner, so the level $s$ is {\em corner-weak}
  against the level $s^\prime$. The levels $s<{8 \over 13}$ are never
  corner-weak against any of the lower level.
  
\item $P_2<P_3$, but $P_1>P_3$, {\it i.e.}  ${1\over
    2}(3-4s)<s^\prime<2-{9\over 4} s$. In this case corner will be stable, so
  the level $s$ will be called {\em corner-stable} against the level
  $s^\prime$.
\end{itemize}

{\bf Step stability}: Consider now the case in which a domain of level
$s^\prime$ forms a step in a domain of level $s$, and $s>s^\prime$. The
payoffs of the three agents marked in Fig. \ref{ECS} are, $P_1=(2-s)s$,
$P_2={1\over 9}(2-s)(8s+s^\prime)$, and $P_3={1\over
  9}(2-s^\prime)(4s+5s^\prime)$. Again there are three possibilities:

\begin{itemize}
\item $P_2>P_3$, {\it i.e.} $s^\prime < {8\over 5}(1-s)$. In this case the
  agent $3$ will be taken over by the agent $2$. Thus the domain of level $s$
  will grow along the step. The level $s$ will be called {\em step-strong}
  against the level $s^\prime$. The levels $s < {8\over 13}$ are step-strong
  against all the lower levels.
  
\item $P_1<P_3$, {\it i.e.}  $2-{9\over 5} s<s^\prime<s$. The domain of
  $s^\prime$ will grow, so the level $s$ is {\em step-weak} against the level
  $s^\prime$. The levels $s<{5 \over 7}$ are never step-weak against any of
  the lower level.
  
\item $P_2<P_3$, but $P_1>P_3$, {\it i.e.}  ${1\over
    2}(3-4s)<s^\prime<2-{9\over 4} s$. In this case step will be stable, so
  the level $s$ will be called {\em step-stable} against the level $s^\prime$.
\end{itemize}

From this simplistic analysis of corner edge and step stability, one can
understand the density profile shown in Fig. \ref{cprofile}.  

The levels $s<{1\over 2}$ do not have any threat from the lower levels, but
they are edge-weak and corner-weak against the higher levels. Therfore they
cannot form colonies and survive just as thin lines.

The levels ${1\over 2} <s < {8\over 13}$ are not edge-weak or corner-weak
against any level.  Therefore they form colonies in the initial stage.
However, because of their step-weakness, the colonies are slowly decimated by
the higher levels.

The level $s={8\over 13}$ is not weak against any level. Therefore, in case of
moderate mutation rate there is a huge jump in cumulative plot at this level
(see Fig. \ref{cprofile}). It constitutes $71\%$  of the whole
population.

The levels ${8\over 13}<s<{5\over 7}$ are not edge-weak or step-weak against
any level. But they are corner-weak against some of the lower levels. In case
of moderate mutation rate these levels are decimated by the level $s={8\over
  13}$ (so there is a huge peak at $s={8\over 13}$). However, in case of very
slow mutation rate these levels can form some colonies by forming a protective
coat of defectors near the corners. Therefore, in this case the population of
cooperators is evenly distributed in the whole range (see Fig.
\ref{cprofile}, the second step is broadened).

The levels $s>{5\over 7}$ are very weak against the lower levels. They are not
even stable as thin lines. Therefore they are completely wiped out from the
population. Their high cooperative tendency works against their survival.

\begin{figure}
  \centering
  \includegraphics[width=0.7\textwidth]{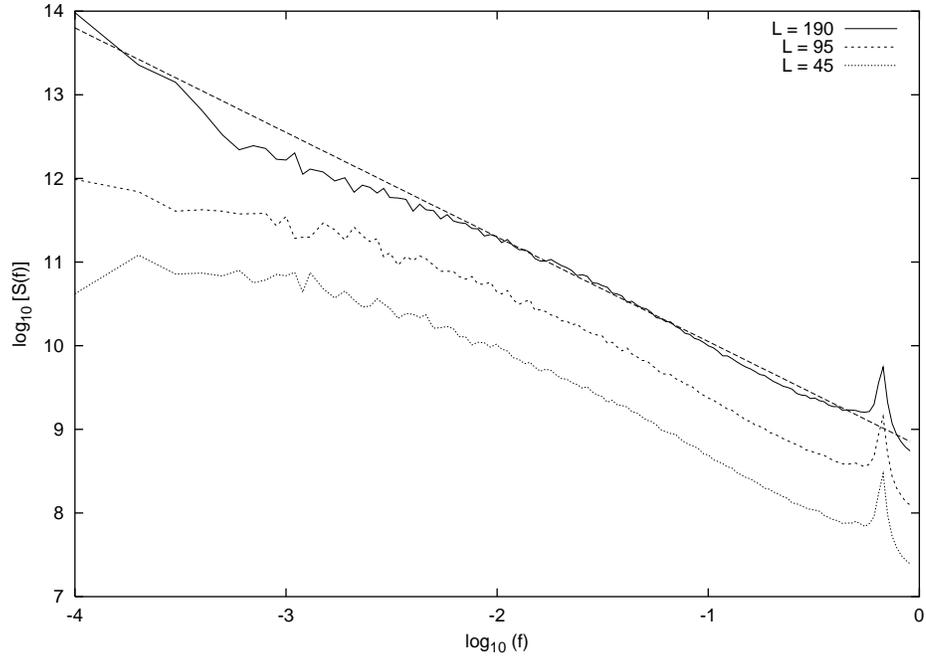}
\caption{Log-log plot of the power spectrum of activity for the lattice size
  $L=45, 95$, and $190$ and a mutation rate $p=0.001$ in case of fixed
  boundary condition.}
\label{sf}
\end{figure}

\begin{figure}
  \centering
  \includegraphics[width=0.7\textwidth]{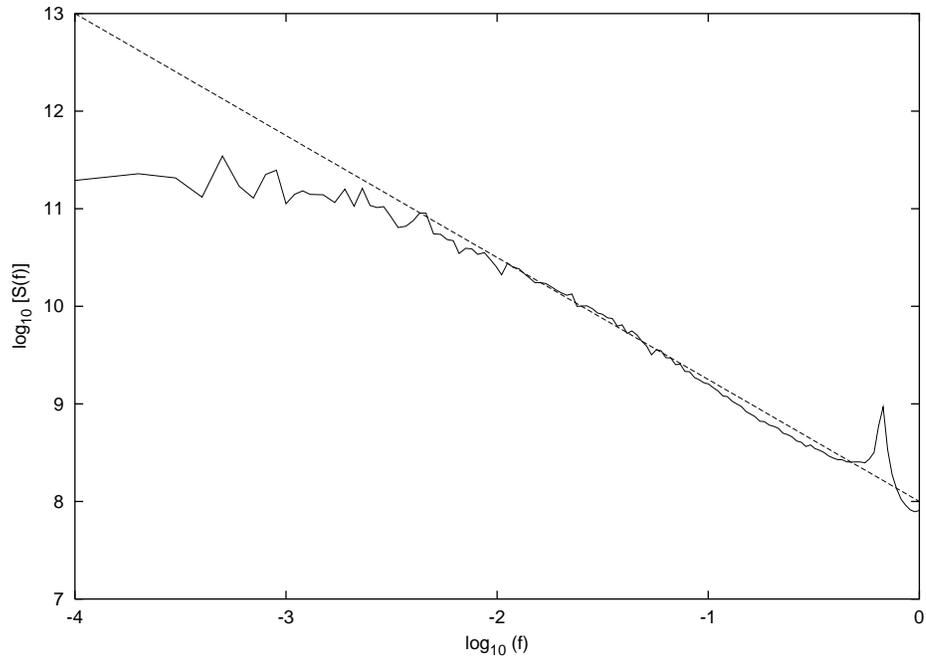}
\caption{Log-log plot of the power spectrum of activity for the lattice size
  $L=95$ and a mutation rate $p=0.001$ in case of periodic boundary
  condition.}
\label{psf}
\end{figure}

\section{Moderate mutation rate}

The behavior of this model depends on the mutation rates. For moderate
mutation rate, i.e., $p > 0.0005$ approximately, the model shows critical
fluctuation in time.  We study the power spectrum of the number of active
sites (sites changing their levels) with both fixed and periodic boundary
conditions. We find that the power spectrum does not depend on the boundary
condition and it shows $1/f^\alpha$ behavior, with $\alpha \sim 1.30$ (see
Fig. \ref{sf}, and \ref{psf}).

The mutation of the occupant of a site creates an avalanche. In case of the
moderate mutation rate, the system does not get enough time to relax, and
these avalanches overlap with each other leading to a bigger avalanche. The
overlap of these avalanches cannot be regarded as uncorrelated superposition
of smaller avalanches because of the temporal correlation. These big
avalanches are responsible for the nontrivial behavior of the power spectrum.
In case of the slow mutation rate, i.e., $p < 0.0005$, the system gets enough
time to relax, therefore the avalanches do not overlap. In this case, the
power spectrum has $1/f^2$ behavior. We note that self-organized criticality
observed in this model, in case of moderate mutation rate, is different from
that of the sandpile model, in which the avalanches do not overlap because of
the slow drive rate \cite{SOC}.

The ``$1/f$'' form of the power spectrum alone does not imply that the model
has long range temporal correlation. This kind of behavior is seen in models
with no translational invariance \cite{RZ,A}.  In this case the
autocorrelation may have exponential form with different relaxation time at
different point in space.  So the power spectrum at each point in space has
the form $1/f^2$, i.e., the power spectrum of a random walk. Averaging over
space, one obtains a power spectrum of form $1/f^\alpha$, with a nontrivial
value of $\alpha$, which is less than $2$.  By this mechanism even a
noncritical models can show a $1/f$ noise \cite{RZ}. Thus $0<\alpha<2$, does
not imply long rang autocorrelation.

\begin{figure}
  \centering
  \includegraphics[width=0.7\textwidth]{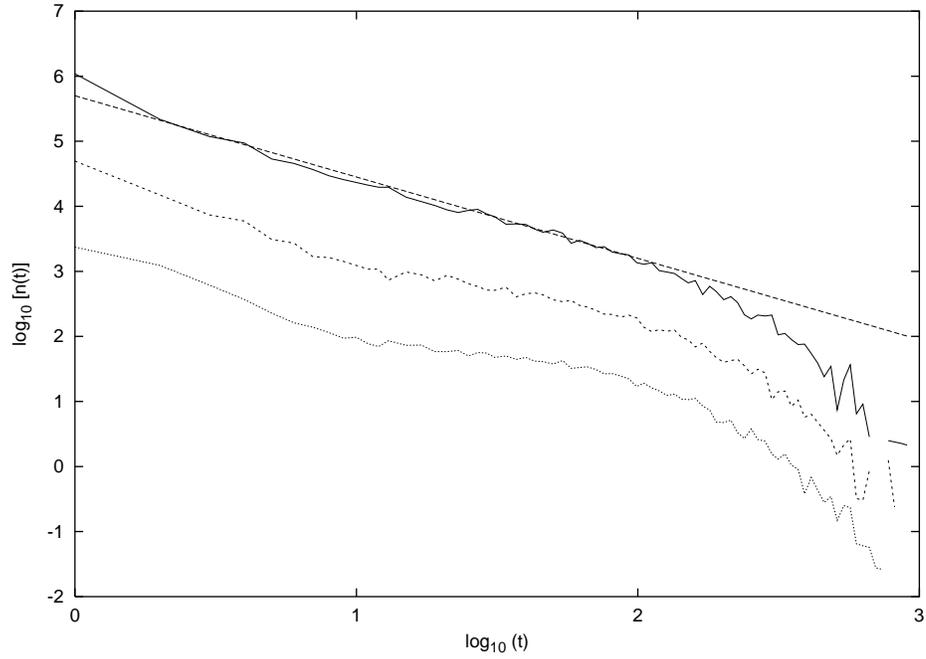}
\caption{Log-log plot of the waiting time distribution for (i)
  a site near the center (top), (ii) a site near the boundary but away from
  corner (middle), and (iii) a site near the corner (bottom). The top and the
  middle graphs have been shifted up by multiplying by 100 and 10
  respectively.}
\label{wt}
\end{figure}

\begin{figure}
  \centering
  \includegraphics[width=0.7\textwidth]{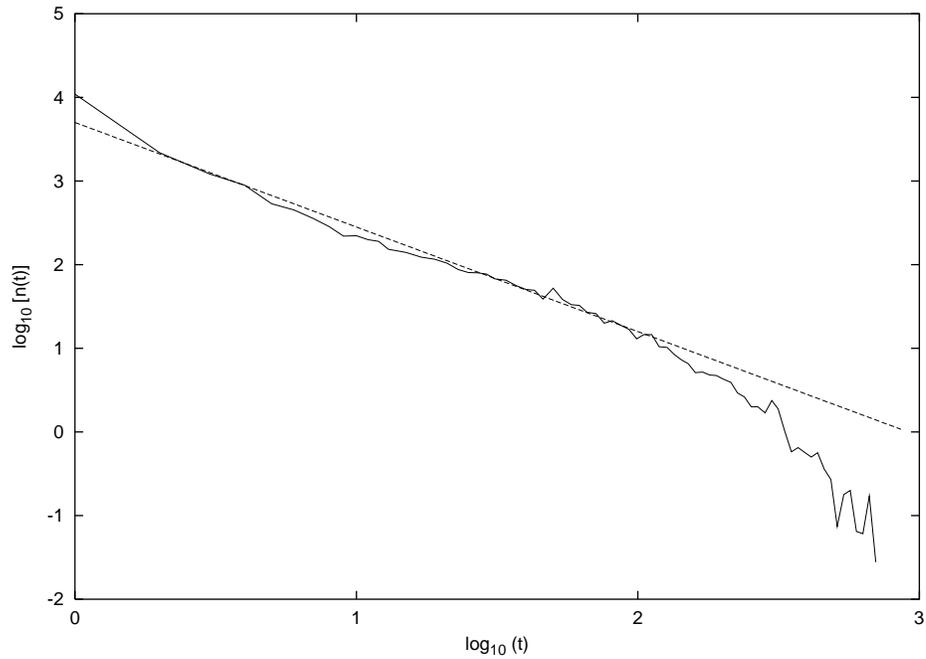}
\caption{Log-log plot of the waiting time distribution for periodic
  boundary condition.}  
\label{pwt}
\end{figure}

In our model there is a spatial pattern (Fig. \ref{cfs}) which could lead to
spatially varying relaxation time. However, this pattern slowly changes in
time because of the random mutations.  To check for the spatially varying
relaxation time, we study the {\em waiting time} (i.e., the time taken between
two consecutive changes of level), with both types of boundary conditions. For
fixed boundary condition we select three different sites on lattice.  The
first site is near the center, the second is near a boundary but away from a
corner, and the third is near the a corner.  The frequency distribution of the
waiting time in the first two cases show a fairly good power law behavior in
two decades (see Fig.  \ref{wt}). It can be fitted to the form: $n(t)\sim
t^{-\beta}$, with $\beta = 1.25$. At the site near a corner, the waiting time
is large and does not fit well with a power law. However, the contribution to
the power spectrum from such a sites is negligible. For periodic boundary
condition, the waiting time distribution is similar to the one obtained for
the site in the bulk with fixed boundary condition (Fig. \ref{pwt}). Thus in
both cases, the relaxation time does not significantly depend on the position.
Thus the ``$1/f$'' behavior of the power spectrum is due to the temporal
correlation of the steady state dynamics.

We have found similar behavior in the power spectrum of the average cooperative
level $\langle S \rangle$.  Thus the exponent $\alpha$ of the power spectrum
appears to be  universally in this model.

Our model has many periodically changing subconfigurations in the steady
state. For this reason, analysis in terms of avalanches produced by small
local perturbations is forbiddingly difficult. Therefore we look for a
different method to investigate the dynamical behavior. The analysis of the
waiting time gives an alternative way to explore the critical dynamics of the
model.

\section{Slow mutation rate}

In case of a slow mutation rate, i.e., $p<0.0005$ approximately, the model
slowly evolves colonies of higher cooperation level, i.e., $s>8/13$ (see Fig.
\ref{dprofile} and \ref{cprofile}. The defectors play an important role in
protecting these colonies against the raid of the level $s=8/13$.  The concave
part (corner) in the boundary of the colony of higher cooperators is weak
against the immediate lower levels.  However, one defector at the corner makes
the colony of higher cooperators stable.

The higher cooperative levels are not able to grow in the initial stage. They
are mainly threatened by the immediate lower levels, rather than the
defectors.  Slowly a small domain of higher level is formed by mutation near
the boundary of the colony of lower cooperators.  This process is normally
very slow and may take hundreds of time steps. The defectors in the
neighborhood of these cooperators become strong and may advance toward the
colony of $s=8/13$, whenever there is a fluctuation at the boundary because of
mutation. Behind the line of these defectors the colony of higher cooperators
grows. The growth is a very slow process and takes tens of thousands of time
steps.

The steady state in this case is not critical. The power spectrum of
fluctuation of activity has a $1/f^2$ form, which is same as the power
spectrum of a random walk.

\section{Summary}
We have explored the possibility of survival of cooperative tendency in a
model of locally shared common resources. The spatial extension of the model
allows the cooperators to form stable colonies (domains). The selfish agents
can only survive in the neighborhood of cooperators, therefore they are found
on the boundaries between these colonies. Thus, natural selection has
provision for coexistence of both type of behaviors.

In addition, this model shows interesting statistical behavior. A moderate
mutation rate makes the model critical with fluctuation of all time scales.
The critical fluctuations are helpful in evolution of cooperative behavior in
the initial stage. However, it work against the evolution of higher
cooperative level. In case of slow mutation rate the system becomes
noncritical and an interesting process of evolution starts. The higher
cooperators form a coat of defectors and slowly advance in the territory of
lower cooperators. Thus the average cooperative level slowly increases.

\end{document}